\documentclass[usenatbib]{mn2e}
\usepackage{graphicx}
\usepackage{times}
\usepackage{amssymb}
\usepackage{natbib}
\def\gsim { \lower .75ex \hbox{$\sim$} \llap{\raise .27ex \hbox{$>$}} }
\def\lsim { \lower .75ex \hbox{$\sim$} \llap{\raise .27ex \hbox{$<$}} }

\begin{document}

\title[Satellites of Simulated Galaxies] 
{Satellites of Simulated Galaxies: survival, merging, and their relation to the dark and
stellar halos}

\author[Sales, Navarro, Abadi \& Steinmetz]{Laura V. Sales$^{1,2}$, Julio F. Navarro,$^{3,4}$\thanks{Fellow of the Canadian Institute for Advanced Research.} Mario G. Abadi $^{1,2,3}$ 
and Matthias Steinmetz$^{5}$
\\
$^{1}$ Observatorio Astron\'omico, Universidad Nacional de C\'ordoba, Laprida
854, 5000 C\'ordoba, Argentina.
\\
$^{2}$ Instituto de Astronom\'{\i}a Te\'orica y Experimental, Conicet, Argentina.
\\
$^{3}$Department of Physics and Astronomy, University of Victoria, Victoria, BC V8P 5C2,
Canada\\
$^{4}$Max-Planck Institut f\"ur Astrophysik, Karl-Schwarzschild Strasse 1,
Garching, D-85741, Germany\\
$^{5}$Astrophysikalisches Institut Potsdam, An der Sternwarte 16, Potsdam 14482, Germany\\
}

\maketitle

\begin{abstract}
We study the population of satellite galaxies formed in a suite of
N-body/gasdynamical simulations of galaxy formation in a $\Lambda$CDM
universe. The simulations resolve the $\sim 10$ most luminous
satellites around each host, and probe systems up to six or seven
magnitudes fainter than the primary. We find little spatial or
kinematic bias between the dark matter and the satellite
population. The radius containing half of all satellites is comparable
to the half-mass radius of the dark matter component, and the velocity
dispersion of the satellites is a good indicator of the virial
velocity of the halo; $\sigma_{\rm sat}/V_{\rm vir} \sim 0.9 \pm 0.2$.
Applied to the Local Group, this result suggests that the virial
velocity of the Milky Way and M31 might be substantially lower than
the rotation speed of their disk components; we find $V_{\rm vir}^{\rm
MW} \sim 109\pm 22$ km/s and $V_{\rm vir}^{\rm M31} \sim 138 \pm 35$
km/s, respectively, compared with $V_{\rm rot}^{\rm MW} \sim 220$ km/s
and $V_{\rm rot}^{\rm M31} \sim 260$ km/s. Although the uncertainties
are large, it is intriguing that both estimates are significantly
lower than expected from some semianalytic models, which predict
a smaller difference between $V_{\rm vir}$ and $V_{\rm rot}$. The
detailed kinematics of simulated satellites and dark matter are also
in good agreement: both components show a steadily decreasing velocity
dispersion profile and a mild radial anisotropy in their velocity
distribution. By contrast, the stellar halo of the simulated galaxies,
which consists predominantly of stellar debris from {\it disrupted}
satellites, is kinematically and spatially distinct from the
population of {\it surviving} satellites. This is because the survival
of a satellite as a self-bound entity depends sensitively on mass and
on time of accretion; surviving satellites are significantly biased
toward low-mass systems that have been recently accreted by the
galaxy. Our results support recent proposals for the origin of the
systematic differences between stars in the Galactic halo and in
Galactic satellites: the elusive ``building blocks'' of the Milky Way
stellar halo were on average more massive, and were accreted (and
disrupted) earlier than the population of dwarfs that has survived
self-bound until the present.
\end{abstract}


\section{Introduction}
\label{sec:intro}

The satellite companions of bright galaxies are exceptionally useful
probes of the process of galaxy formation. Studies of the dynamics of
satellites around bright galaxies, for example, have provided
incontrovertible evidence for the ubiquitous presence of massive dark
halos surrounding luminous galaxies, a cornerstone of the present
galaxy formation paradigm.  Following the pioneering work of 
\cite{holmberg69,zaritsky93,zaritsky97a} compiled perhaps the first
statistically-sound sample of satellite-primary systems with accurate
kinematics, and were able to provide persuasive evidence that the dark
matter halos hinted at by the flat rotation curves of spiral galaxies
(\citealt{sofue01} and references therein) truly dwarf the mass of
the luminous component and extend well beyond the luminous radius of
the central galaxy.

Satellite dynamical studies have entered a new realm since the advent
of large redshift surveys, such as the 2dfGRS \citep{colless01} and
the SDSS \citep{york00,strauss02}, which have increased
many-fold the number of primary-satellite systems known. Recent work
based on these datasets have corroborated and extended the results of
Zaritsky et al, and their conclusions now appear secure. The dynamics
of satellites confirm (i) that dark matter halos extend to large
radii, (ii) that more massive halos surround brighter galaxies, and
(iii) that early-type galaxies are surrounded by halos about twice as
massive as late-type systems of similar luminosity (\citealt{mckay02,
prada03,brainerd04a, vandenbosch05}; see
\citealt{brainerd04b} for a recent review).

Satellites may also be thought of as probes of the faint end of the
luminosity function. After all, satellite galaxies are, by definition,
dwarf systems, thought to be themselves surrounded by their own
low-mass dark matter halos. These low-mass halos are expected to be
the sites where the astrophysical processes that regulate galaxy
formation (i.e., feedback) operate at maximum efficiency. Thus, the
internal structure, star formation history, and chemical enrichment of
satellites provide important constraints on the process of galaxy
formation in systems where theoretical models predict a highly
non-trivial relation between dark mass and luminosity (see, e.g.,
\citealt{whiteandrees78, kauffmann93, cole94}; see as well 
\citealt{cole00} and \citealt{benson02} for a more
detailed list of references).

The anticipated highly non-linear mapping between dark matter and
light at the faint-end of the luminosity function is perhaps best
appreciated in the satellite population of the Local Group, where the
relatively few known satellites stand in contrast with the {\it
hundreds} of ``substructure'' cold dark matter (CDM) halos of
comparable mass found in cosmological N-body simulations
(\citealt{klypin99b,moore99}). Possible resolutions of this ``satellite
crisis'' have been discussed by a number of authors, and there is
reasonably broad consensus that it originates from inefficiencies in
star formation caused by the combined effects of energetic feedback
from evolving stars and by the diminished supply of cold gas due to
reionization \citep[see, e.g.][]{kauffmann93, bullock00, 
somerville01, benson02}. These effects combine to reduce
dramatically the star formation activity in substructure halos, and
can reconcile, under plausible assumptions, the substructure halo mass
function with the faint end of the satellite luminosity function
\citep{stoehr02, kazantzidis04, penarrubia07}.

The price paid for reconciling cold dark matter substructure with the
Local Group satellite population is one of simplicity, as the
``feedback'' processes invoked involve complex astrophysics that is
not yet well understood nor constrained. It is not yet clear, for
example, whether the brighter satellites inhabit the more massive
substructures, or whether, in fact, there is even a monotonic relation
between light and mass amongst satellites. This issue is further
complicated by the possibility that a substantial fraction of a
satellite's mass may have been lost to tides. Tidal stripping is
expected to affect stars and dark matter differently, complicating
further the detailed relation between light and mass in substructure
halos (\citealt{hayashi03, kravtsov04}, Strigari et al. 2007a,b).
\nocite{strigari07a,strigari07b}

These uncertainties hinder as well the interpretation of satellites as
relics of the hierarchical galaxy assembly process, and consensus has
yet to emerge regarding the severity of the biases that the various
effects mentioned above may engender.  Do the spatial distribution of
satellites follow the dark matter? Is the kinematics of the satellite
population substantially biased relative to the dark matter's? Have
satellites lost a substantial fraction of their stars/dark matter to
stripping? Are surviving satellites fair tracers of the population of
accreted dwarfs? 

Of particular interest is whether satellites may be considered relics
of the ``building blocks'' that coalesced to form the early
Galaxy. Indeed, the stellar halo of the Milky Way is regarded, in
hierarchical models, to consist of the overlap of the debris of many
accreted satellites which have now merged and mixed to form a
kinematically hot, monolithic stellar spheroid \citep{searleandzinn78,
bullockandjohnston05, abadi06, moore06} . A
challenge to this view comes from detailed observation of stellar
abundance patterns in satellite galaxies in the vicinity of the Milky
Way. At given metallicity, the stellar halo (at least as sampled by
stars in the solar neighbourhood) is systematically more enriched in
$\alpha$-elements than stars in Galactic satellites \citep{fuhrmann98,
shetrone01, shetrone03, venn04}, a result that remains
true even when attempting to match stars of various ages or
metallicities \citep{unavane96, gilmoreandwyse98, pritzl05}. 
Can hierarchical models explain why satellites identified today
around the Milky Way differ from the ones that fused to form the
Galactic halo?

Preliminary clues to these questions have been provided by the
semianalytic approach of Bullock, Johnston and collaborators 
\citep{bullockandjohnston05, font06a, font06b}, who argue that hierarchical
models predict naturally well-defined distinctions between the halo
and satellite stellar populations. Detailed answers, however, depend
critically on which and when substructure halos are ``lit up'' and how
they evolve within ``live'' dark matter halos. These are perhaps best
addressed with direct numerical simulation that incorporates the
proper cosmological context of accretion as well as the gasdynamical
effects of cooling and star formation in an evolving population of
dark matter halos. The study we present here aims to address these
issues by analyzing the properties of the satellite population of
$L_*$ galaxies simulated in the $\Lambda$CDM scenario. We introduce
briefly the simulations in \S~\ref{sec:numexp}, analyze and discuss
them in \S~\ref{sec:analysis} and we conclude with a summary in
\S~\ref{sec:conc}.

\begin{figure*}
\begin{center}
\includegraphics[width=\linewidth,clip]{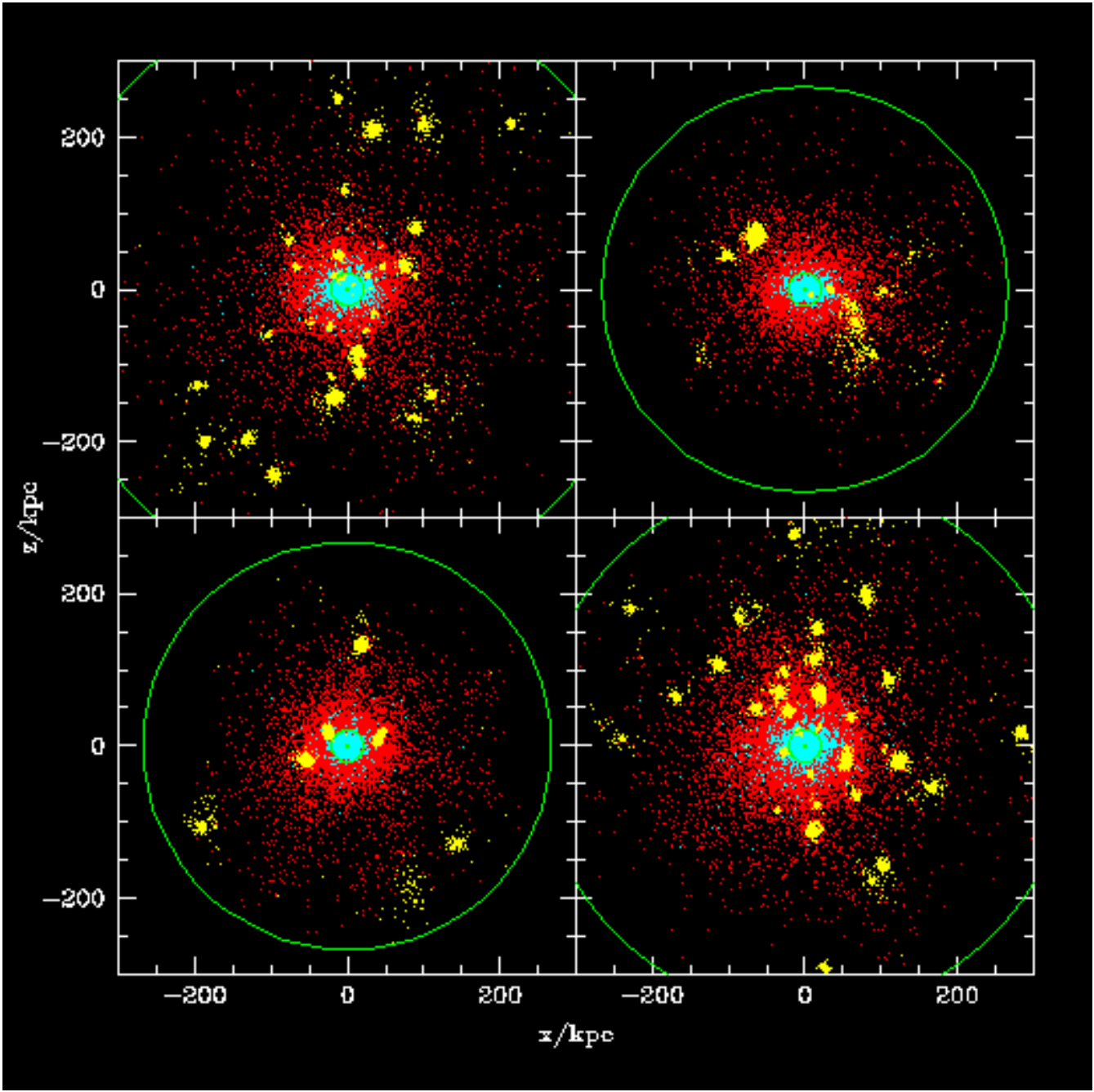}
\end{center}
\caption{Spatial distribution of the stellar component of four of our
  simulated galaxies at z=0. Each panel corresponds to a different
  simulation, projected so that the inner galaxy is seen approximately
  ``edge-on''. The virial radius of the system is marked by the outer
  green circle in each panel. The inner circle has a radius of $20$
  kpc, where most the stars in each galaxy are found. Stars that have
  formed in satellites that survive as self-bound entities until $z=0$
  are shown in yellow. ``In situ'' stars, i.e., those formed in the
  most massive progenitor of the galaxy, are shown in cyan, whereas
  those formed in satellites that have been accreted and disrupted by
  the main galaxy are shown in red. Note that the diffuse outer
  stellar halo reaches almost out to the virial radius, and consists
  almost exclusively of accreted stars. The inner galaxy, on the other
  hand, is dominated by stars formed ``in situ''.
\label{fig:xypanel}}
\end{figure*}

\section{The Numerical Simulations}
\label{sec:numexp}

Our suite of eight simulations of the formation of $L_*$ galaxies in
the $\Lambda$CDM scenario is the same discussed recently by 
Abadi, Navarro and Steinmetz 2006. The simulations are similar to the one
originally presented by \citet{steinmetzandnavarro02}, and have been
analyzed in detail in several recent papers, which the interested
reader may wish to consult for details on the numerical setup 
\citep{abadi03a, abadi03b, meza03, meza05, navarro04}. 

Briefly, each simulation follows the evolution of a small region of
the universe chosen so as to encompass the mass of an $L_{*}$ galaxy
system. This region is chosen from a large periodic box and
resimulated at higher resolution preserving the tidal fields from the
whole box. The simulation includes the gravitational effects of dark
matter, gas and stars, and follows the hydrodynamical evolution of the
gaseous component using the Smooth Particle Hydrodynamics (SPH)
technique \citep{steinmetz96}. We adopt the following cosmological
parameters for the $\Lambda$CDM scenario: $H_0=65$ km/s/Mpc,
$\sigma_8=0.9$, $\Omega_{\Lambda}=0.7$, $\Omega_{\rm CDM}=0.255$,
$\Omega_{\rm bar}=0.045$, with no tilt in the primordial power
spectrum. 

All simulations start at redshift $z_{\rm init}=50$, have force
resolution of order $1$ kpc, and the mass resolution is chosen so that
each galaxy is represented on average, at $z=0$, with $\sim 50,000$
gas/dark matter particles and $\sim 125,000$ star particles. Each
re-simulation follows a single $\sim L_*$ galaxy in detail, and
resolves a number of smaller, self-bound systems we shall call
generically ``satellites''. We shall hereafter refer to the main
galaxy indistinctly as ``primary'' or ``host''.

Gas is allowed to turn into stars at rates consistent with the
empirical Schmidt-like law of \citet{kennicutt98} in collapsed regions at
the center of dark matter halos.  Because star formation proceeds
efficiently only in high-density regions, the stellar components of
primary and satellites are strongly segregated spatially from the dark
matter. We include the energetic feedback of evolving stars, although
its implementation mainly as a heating term on the (dense) gas
surrounding regions of active star formation implies that most of this
energy is lost to radiation and that feedback is ineffective at
curtailing star formation.  The transformation of gas into stars thus
tracks closely the rate at which gas cools and condenses at the center
of dark matter halos.  This results in an early onset of star-forming
activity in the many progenitors of the galaxy that collapse at high
redshift, as well as in many of the satellite systems we analyze here.

Another consequence of our inefficient feedback algorithm is that gas
cooling and, therefore, star formation, proceed with similar
efficiency in all well-resolved dark matter halos, irrespective of
their mass. As a result, the total stellar mass of a satellite
correlates quite well with the ``original'' mass of its progenitor
dark halo; i.e., the total mass of the satellite before its accretion
into the virial radius of its host. We define the {\it virial} radius,
$r_{\rm vir}$, of a system as the radius of a sphere of mean density
$\Delta_{\rm vir}(z)$ times the critical density for closure. This
definition defines implicitly the virial mass, $M_{\rm vir}$, as that
enclosed within $r_{\rm vir}$, and the virial velocity, $V_{\rm vir}$,
as the circular velocity measured at $r_{\rm vir}$. Quantities
characterizing a system will be measured within $r_{\rm vir}$, unless
otherwise specified. The virial density contrast, $\Delta_{\rm
vir}(z)$ is given by $\Delta_{\rm vir}(z)=18\pi^2+82f(z)-39f(z)^2$,
where $f(z)=[\Omega_0(1+z)^3/(\Omega_0(1+z)^3+\Omega_\Lambda))]-1$ and
$\Omega_0=\Omega_{\rm CDM}+\Omega_{\rm
bar}$\citep{bryanandnorman98}. $\Delta_{\rm vir}\sim 100$ at $z=0$.

It is likely that improvements to our feedback algorithms may lead to
revisions in the efficiency and timing of star formation in these
galaxies, and especially in the satellites, but we think our results
will nonetheless apply provided that these revisions do not
compromise the relatively simple relation between stellar mass and
halo mass that underpins many of our results. For example, we expect
that modifications of the star formation algorithm will affect
principally the number, age, and chemical composition of stars, rather
than the dynamical properties of the satellites. This is because the
latter depend mainly on the mass, orbit, and timing of the merging
progenitors, which are largely dictated by the assumed cosmological
model. These properties are less sensitive to the complex astrophysics
of star formation and feedback, and therefore our analysis focuses on
the kinematics and dynamical evolution of the satellite population
around the eight galaxies in our simulation suite.

\begin{center}
\begin{figure}
\includegraphics[width=84mm]{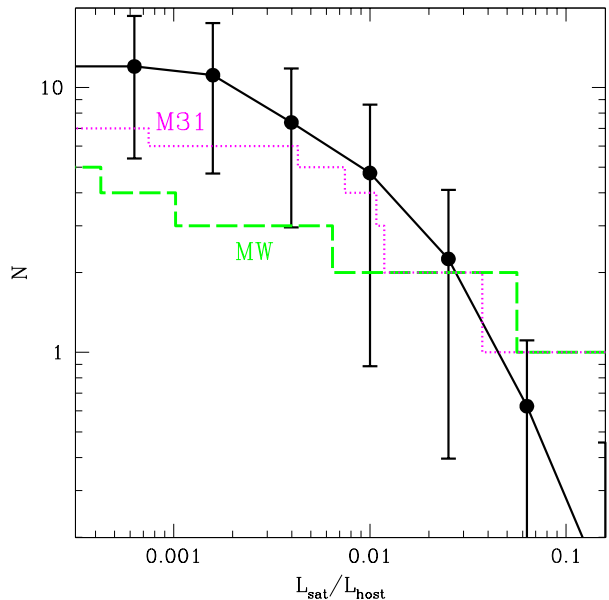}
\caption{Cumulative luminosity distribution of simulated satellites
(filled circles), averaged over our eight simulations, and compared
with the Milky Way (blue dashed line) and M31 (red dotted curve)
satellite systems. Satellite luminosities are scaled to the luminosity
of the host. Error bars in the simulated data indicate Poisson
uncertainties in the computation of the average. The flattening of the
simulated satellite distribution below $0.1\%$ of the primary
luminosity is due to numerical limitations. The Local Group data are
taken from \citet{vandenbergh99}. For the MW and M31 systems we include
only satellites at distances closer than $300$ kpc from the central
galaxy.}
\label{fig:nlv}
\end{figure}
\end{center}

\begin{center}
\begin{figure}
\includegraphics[width=84mm]{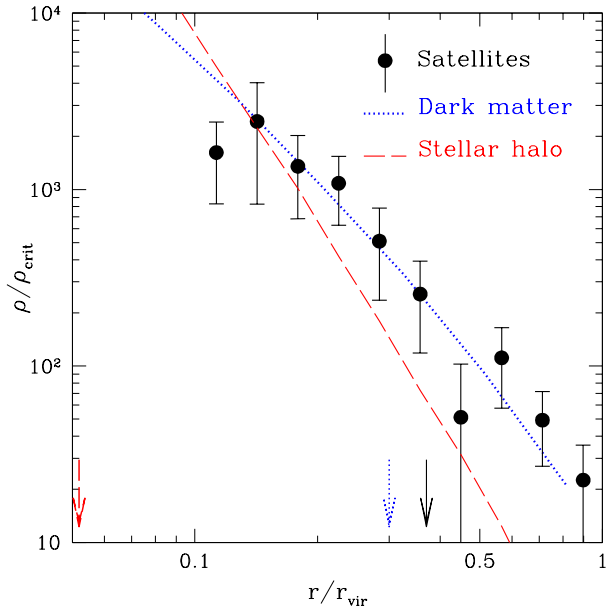}
\caption{Number density profile of simulated satellites, after scaling
their positions to the virial radius of each host and stacking all
eight simulations (solid circles; error bars denote Poisson
uncertainties associated with the total number of satellites in each
radial bin). The dotted line corresponds to the average dark matter
density profile, and the dashed line to the stars in the outer stellar
halo. The vertical normalization for the satellite and stellar halo
profiles has been chosen so that all profiles approximately coincide
at $r\sim 0.15 \, r_{\rm vir}$. Note that the spatial distribution of
satellites is similar to the dark matter, and that stars in the
stellar halo are significantly more centrally concentrated. Arrows
mark the radius containing half the objects in each component. See
text for further discussion.}
\label{fig:ndprof}
\end{figure}
\end{center}

\section{Results and Discussion}
\label{sec:analysis}

\subsection{Characterization of the satellite population}
\label{ssec:char}

Figure~\ref{fig:xypanel} shows the spatial distribution of all star
particles in four of our simulated galaxies. Stars are assigned to one
of three components and colored accordingly. Particles in cyan are
``in-situ'' stars; i.e., stars that formed in the main progenitor of
the primary galaxy. Stars in red formed in satellites that have since
been accreted and fully disrupted by the tidal field of the
galaxy. Stars in yellow formed in systems that survive as recognizable
self-bound satellites until $z=0$. As discussed in detail by Abadi et
al (2006), the tidal debris of fully disrupted satellites makes up the
majority of the smooth outer stellar halo component.  ``In-situ''
stars, on the other hand, dominate the inner galaxy, whereas surviving
satellites are easily identifiable as overdense, tightly bound clumps
of stars.

In practice, we identify satellite systems using a friends-of-friends
algorithm to construct a list of potential stellar groupings, each of
which is checked to make sure that (i) they are self-bound, and that
(ii) they contain at least $35$ star particles. This minimum number of
stars (which corresponds roughly to $\sim 0.03\%$ of the stellar mass
of the primary at $z=0$) is enough to ensure the reliable
identification of the satellite at various times and the robust
measurement of their orbital properties, but is insufficient to study
the internal structure of the satellite. The satellite identification
procedure is run for all snapshots stored for our simulations,
allowing us to track the evolution of individual satellites.

With these constraints, our simulations resolve, at $z=0$, an average
of about $10$ satellites within the virial radius of each simulated
galaxy. The cumulative luminosity distribution of these satellites
(computed in the $V$ band{\footnote{Luminosity estimates in various
bands are made by convolving the masses and ages of star particles
with standard spectrophotometric models, see, e.g., Abadi et al 2006
for details.}} for ease of comparison with data available for the Local
Group satellites) is shown in Figure~\ref{fig:nlv}. The brightest
satellite is, on average, about $\sim 12\%$ as bright as the primary,
in reasonable agreement with the most luminous satellite around the
Milky Way and M31: the LMC and M33 are, respectively, $11\%$ and
$8\%$ as bright as the MW and M31 (van den Bergh 1999).

At brightnesses below $0.2\%$ of $L_{\rm host}$ the number of
simulated satellites levels off as a result of numerical
limitations. Independent tests (Abadi et al, in preparation) show that
this brightness limit corresponds to where satellite identification in
the simulations becomes severely incomplete. We note that this
limitation precludes us from addressing the ``satellite crisis''
alluded to in \S1: our simulations lack the numerical resolution
needed to resolve the hundreds of low-mass substructure halos found in
higher-resolution CDM simulations. On average, the $10$th brightest
satellite in our simulations is $\sim 5.6$ mag fainter than the
primary; for comparison, the MW and M31 have only $2$ and $5$
satellites as bright as that. 

Given the small number of systems involved and the considerable
scatter from simulation to simulation (the number of bright satellites
ranges from $4$ to $21$ in our eight simulations) we conclude that
there is no dramatic discrepancy between observations and simulations
at the bright end of the satellite luminosity function. Applying our
results to the full Local Group satellite population, including, in
particular, the extremely faint dwarfs being discovered by panoramic
surveys of M31 and by the SDSS (\citealt{zucker04,zucker06,willman05b,
martin06,belokurov06,belokurov07,irwin07,majewski07}, Ibata et al. 2007
submitted), involves a 
fairly large extrapolation, and should be undertaken with 
caution (see, e.g., Pe\~narrubia, McConnachie \& Navarro 2007 
for a recent discussion).

\begin{center}
\begin{figure}
\includegraphics[width=84mm]{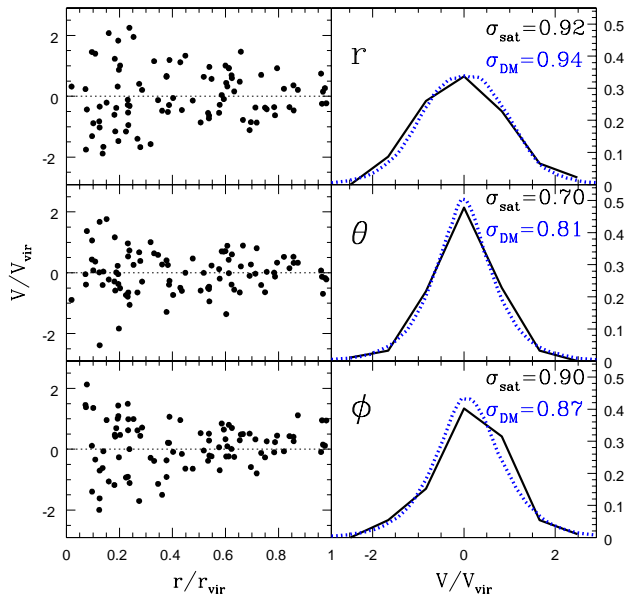}
\caption{Spherical components of satellite velocities at $z=0$ as a
function of their distance to the center of the host galaxy. Each
system has been rotated so that the angular momentum of the inner
galaxy is aligned with the direction of the $z$-coordinate
axis. Positions and velocities have been scaled to the virial radius
and velocity of each host halo. Panels on the right show the velocity
distributions of the satellite population within $r_{\rm vir}$ (solid
lines) and compare it to the dark matter particles (dotted lines).
The velocity dispersions are given in each panel. Note the slight
asymmetry in the satellites' $V_{\phi}$ velocity distribution, which
results from the net co-rotation of satellites around the primary.}
\label{fig:vrtp}
\end{figure}
\end{center}

\begin{center}
\begin{figure}
\includegraphics[width=84mm]{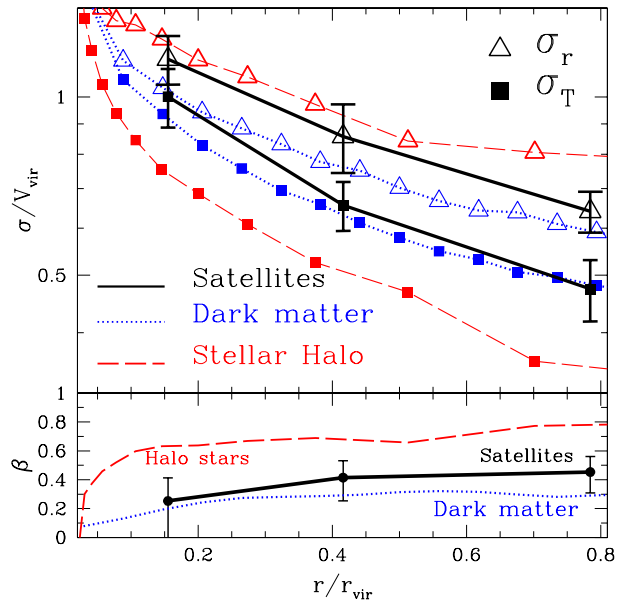}
\caption{{\it Top panel:} Radial and tangential velocity dispersion
profiles of satellites, dark matter, and stellar halo, computed after
scaling to virial values and stacking all simulations in our
series. {\it Bottom panel:} Anisotropy parameter as a function of
radius for the satellite population, compared with dark matter
particles and with the stellar halo. Note that satellites are only
slightly more radially anisotropic than the dark matter and
kinematically distinct from the stellar halo.}
\label{fig:rdisp}
\end{figure}
\end{center}

\subsection{Spatial distribution}
\label{ssec:satvr}

Figure~\ref{fig:xypanel} shows that satellites are found throughout
the virial radius of the host and that, unlike stars in the smooth
stellar halo, satellites show little obvious preference for clustering
in the vicinity of the central galaxy. This is confirmed in
Figure~\ref{fig:ndprof}, where the solid circles show the number
density profile of satellites, after rescaling their positions to the
virial radius of each host and stacking all eight simulations. The
dashed and dotted lines in this figure correspond, respectively, to
the density profile of the stellar and dark matter halos, scaled and
stacked in a similar way. The vertical normalization of the satellite
and stellar halo profiles is arbitrary, and has been chosen so that
all profiles approximately match at $r\sim 0.15 \, r_{\rm vir}$.

There is little difference in the shape of the dark matter and
satellite profiles: half of the satellites are contained within $\sim
0.37 \, r_{\rm vir}$, a radius similar to the half-mass radius of the
dark matter, $\sim 0.3 \, r_{\rm vir}$. We conclude that, within the
uncertainties, the satellites follow the dark matter. The stellar
halo, on the other hand, is much more centrally concentrated than the
dark matter and satellites; its half-mass radius is only $\sim 0.05 \,
r_{\rm vir}$, as shown by the arrows in Figure~\ref{fig:ndprof}.

This result implies that the spatial distribution of simulated
satellites is distinct from that of CDM substructure halos, whose
density profile is known to be significantly shallower than the dark
matter's \citep{ghigna98, ghigna00, gao04a, diemand04}. This suggests
that the ``mapping'' between dark and luminous substructure is highly
non-trivial, as argued by \citet{springel01} and \citet{delucia04}.
Our results, which are based on direct numerical simulation, validate
these arguments and illustrate the complex relation between galaxies
and the subhalos in which they may reside \citep[see
also][]{kravtsov04, nagaiandkravtsov05, gnedin06, weinberg06, libeskind07}.
Luminous satellites are resilient to disruption by tides, and they can
survive as self-bound entities closer to the primary, where
substructures in dark matter-only simulations may quickly disrupt, as
first pointed out by White \& Rees (1978).

We conclude that using dark matter substructures to trace directly the
properties of luminous satellites is likely to incur substantial and
subtle biases which may be difficult to avoid. Models that attempt to
follow the evolution of dark matter substructures and their luminous
components are likely to fare better \citep[see, e.g.][]{croton06,
bower06}. At the low mass end, the inclusion of some treatment of the
substructure mass loss and tidal shocks is needed to put in better
agreement semianalytic models with the results from numerical simulations
\citep{taylorandbabul01,benson02a}.
Definitive conclusions will probably need to
wait until realistic simulations with enhanced numerical resolution
and improved treatment of star formation become available.

\subsection{Kinematics}

The likeness in the spatial distribution of satellites and dark matter
anticipates a similar result for their kinematics. This is illustrated
in Figure ~\ref{fig:vrtp}, where the panels on the left show the
spherical components of the satellites' velocities (in the rest frame
of the host and scaled to its virial velocity) versus galactocentric
distance (in units of the virial radius of the host). Velocity
components are computed after rotating each system so that the
$z$-axis (the origin of the polar angle $\theta$) coincides with the
rotation axis of the inner galaxy. The corresponding velocity
distributions are shown by the thick solid lines in the panels on the
right, and compared with those corresponding to the dark matter
particles (dotted lines). 

The velocity distribution of each component is reasonably symmetric
and may be well approximated by a Gaussian, except perhaps for the
satellites' $V_{\phi}$-component, which is clearly asymmetric. This is
a result of net rotation around the $z$ axis: the satellite population
has a tendency to co-rotate with the galaxy's inner body which is more
pronounced than the dark matter's. Indeed, we find that on average the
specific angular momentum of satellites is $\sim 50\%$ higher than the
dark matter, and a factor of $\sim 10$ higher than the stellar
halo. This result likely arises as a consequence of the accretion and
survival biases discussed below; surviving satellites accrete late and
from large turnaround radii, making them especially susceptible to the
tidal torques responsible for spinning up the galaxy.  The overall
effect, however, is quite small, and rotation provides a negligible
amount of centrifugal support to the satellite population.

The velocity dispersion of both satellites and dark matter particles
drops steadily from the center outwards, as shown in
Figure~\ref{fig:rdisp}. The top panel shows that the drop is similar in
all components, and that the velocity dispersion decreases from its
central value by a factor of $\sim 2$ at the virial radius. This
figure also shows that the velocity distribution is radially
anisotropic, and that the anisotropy becomes more pronounced in the
outer regions. The trends are again similar for satellites and dark
matter, rising slowly with radius and reaching $\beta\sim 0.4$ at the
virial radius. (The anisotropy parameter, $\beta$, is given by $\beta
= 1-({\sigma_t}^2/2{\sigma_r}^2)$, where $\sigma_r$ is the radial
velocity dispersion and
$\sigma_t=\sqrt{(\sigma_\phi^2+\sigma_\theta^2)/2}$ is the tangential
velocity dispersion.)

The stellar halo, on the other hand, is kinematically distinct from
the satellites and from the dark matter. Overall, its velocity
dispersion is lower, and its anisotropy is more pronounced, rising
steeply from the center outwards and becoming extremely anisotropic
($\beta \sim 0.8$) in the outer regions. As discussed in detail by
Abadi et al (2006), this reflects the origin of the stellar halo as
debris from satellite disruption, which occur at small radii, where
tidal forces are maximal. Stars lost during disruption (merging)
events and that now populate the outer halo must therefore be on
rather eccentric orbits, as witnessed by the prevalence of radial
motions in Figure~\ref{fig:rdisp}. The kinematical distinction between
satellites and stellar halo thus suggests that few halo stars have been
contributed by stripping of satellites that have survived self-bound
until the present. We shall return to this issue below.

\begin{center}
\begin{figure}
\includegraphics[width=84mm]{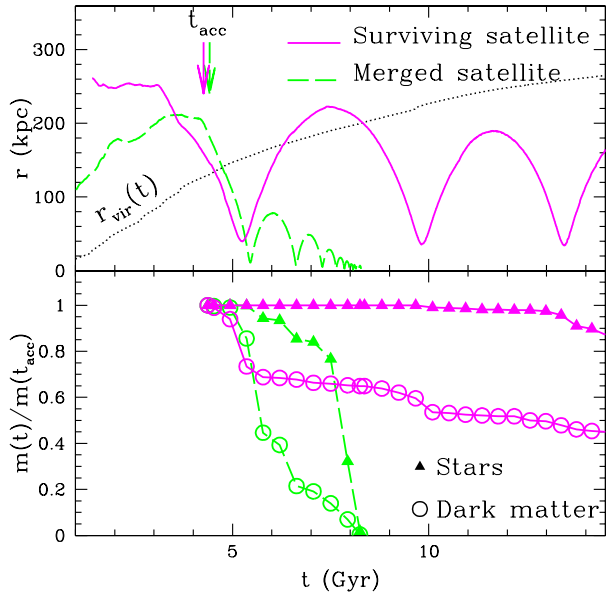}
\caption{{\it Top panel:} Orbital evolution of two satellites, chosen
to illustrate the case of a system that merges quickly with the
primary and of another that survives as a self-bound entity until
$z=0$. Curves show the distance from the primary to the self-bound
stellar core of the satellite as a function of time. The dotted line
shows the evolution of the virial radius of the primary galaxy, and
the arrow indicates the time, $t_{\rm acc}$, when the satellites are
first accreted into the primary's halo. Although both satellites are
accreted more or less at the same time, they are not a physical pair
and evolve independently. {\it Bottom panel:} The evolution of the
satellites' bound mass of stars and dark matter, normalized to the
values computed at the time of accretion. Note that the stellar
component is much more resilient to the effect of tides.}
\label{fig:orbmass}
\end{figure}
\end{center}

\subsection{Application to the Local Group}

The lack of strong kinematical bias between satellites and dark matter
may be used to estimate the virial velocity of the Milky Way and M31.
For example, assuming that the radial velocity dispersion of the
satellites is related to the virial velocity by $\sigma_r \sim 0.9 \,
(\pm 0.2) V_{\rm vir}$ (see Figure~\ref{fig:vrtp}; the uncertainty is
just the rms scatter from our eight simulations), we obtain $V_{\rm
vir}^{\rm MW}\sim 109\pm 22$ km/s from the $\sim 99$ km/s
Galactocentric radial velocity dispersion of the eleven brightest
satellites (see, e.g., the compilation of van den Bergh 1999).

The same procedure may be applied to M31 satellites. Taking into
account projection effects, we find that the line-of-sight satellite
velocity dispersion is $\sigma_{\rm los} \sim 0.8 (\pm 0.2) V_{\rm
vir}$. Taking the $16$ brightest satellites within $300$ kpc from the
center of M31, we find $\sigma_{\rm los} \sim 111$ km/s, implying
$V_{\rm vir}^{\rm M31} \sim 138 \pm 35$ km/s. We use here the
compilation of \citet{mcconnachieandirwin06}, complemented with data
for And XIV from \citet{majewski07}, and for And XII from (Chapman et
al 2007, submitted).

These results imply that the virial radius of the Milky Way is $r_{\rm
vir}^{\rm MW} \sim 240$ kpc. Our simulations predict that half of the
brightest satellites should be enclosed within $\sim 90$ kpc, which
compares favourably with observations: half of the eleven brightest
satellites are within $\sim 90.1$ kpc from the center of the Milky
Way. Contrary to the arguments of \citet{taylor05}, no
substantial bias between satellites and dark matter is required to
explain the MW satellite spatial distribution, provided that one
accepts a virial radius as small as $\sim 240$ kpc.

The same argument, applied to M31, suggests that half of the 16
satellites within its virial radius ($r_{\rm vir}^{\rm M31} \sim 300$
kpc) must be within $\sim 111$ kpc, compared with the observational
value of $\sim 165 $ kpc. Note that these radii are actual distances to M31,
rather than projections. 

Despite the sizable statistical uncertainty inherent to the small
number of satellites in these samples, it is interesting that both of
the virial velocity estimates mentioned above are significantly lower
than the rotation speed measured for these galaxies in the inner
regions; $V_{\rm rot}^{\rm MW}\sim 220$ km/s and $V_{\rm rot}^{\rm
M31} \sim 260 $ km/s.  These low virial velocity estimates are in line
with recent work that advocates relatively low masses for the giant
spirals in the Local Group \citep{klypin02, seigar06,abadi06, smith06}.

If confirmed, this would imply that the circular velocity should drop
steadily with radius in the outer regions of these galaxies. As
discussed by Abadi et al (2006), this may be the result of ``adiabatic
contraction'' of the dark matter halo following the assembly of the
luminous galaxy. However, such result may be difficult to reconcile
with semianalytic models of galaxy formation, which favor a better
match between $V_{\rm rot}$ and $V_{\rm vir}$. Croton et al (2006)
argue that $V_{\rm rot}$ should be similar to the maximum circular
velocity of the dark matter halo, which is only about $\sim 20\%$
larger than $V_{\rm vir}$ for typical concentrations.  It is possible
that taking into account the effects of the adiabatic contraction and
including the self-gravity of the baryon material might induce a large
scatter and allow rotation speeds as high as $V_{\rm rot}\sim 1.5-2$
times $V_{\rm vir}$ (A.Benson, private communication).  Final word on
this issue needs further data to place better constraints on the mass
of the halo of the Local Group spirals at large distances, as well as
improved semianalytic modeling that re-examines critically the
response of the dark halo to the formation of the luminous galaxy. At
least from the observational point of view, the steady pace of
discovery of new satellites of M31 and MW facilitated by digital sky
surveys implies that it should be possible to revisit this issue in
the near future with much improved statistics.

\begin{center}
\begin{figure}
\includegraphics[width=84mm]{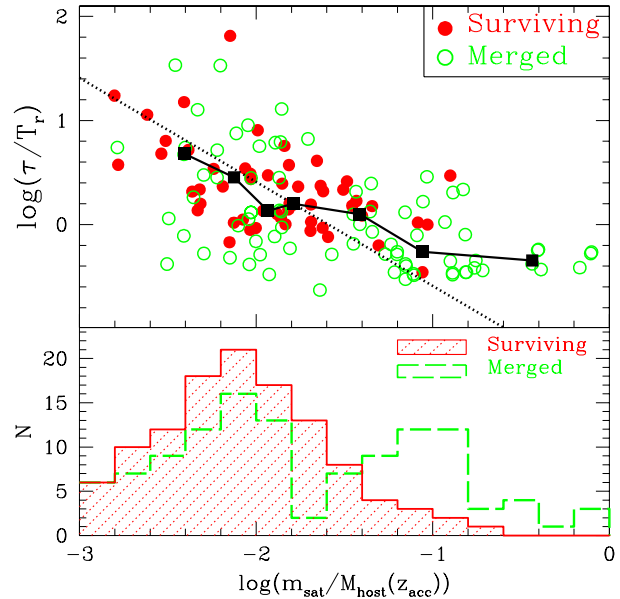}
\caption{{\it Top panel:} Orbital decay timescale of satellites,
$\tau$, shown as a function of satellite mass. Decay timescales are
computed by fitting an exponential law to the evolution of the
apocentric radius of a satellite, and is shown in units of the
(radial) orbital period measured at accretion time. Satellite masses
(dark+baryons) are scaled to the total mass of the host at $t_{\rm
acc}$. Filled and open circles correspond to satellites that have,
respectively, survived or merged with the primary by $z=0$. Filled
squares show the median decay timescale after splitting the sample
into equal-number mass bins. More massive satellites spiral in faster
due to the effects of dynamical friction. {\it
Bottom panel:} Histogram of surviving and merged satellites as a
function of satellite mass. Note the strong mass bias of surviving
satellites relative to merged ones. }
\label{fig:orbdec}
\end{figure}
\end{center}

\begin{center}
\begin{figure}
\includegraphics[width=84mm]{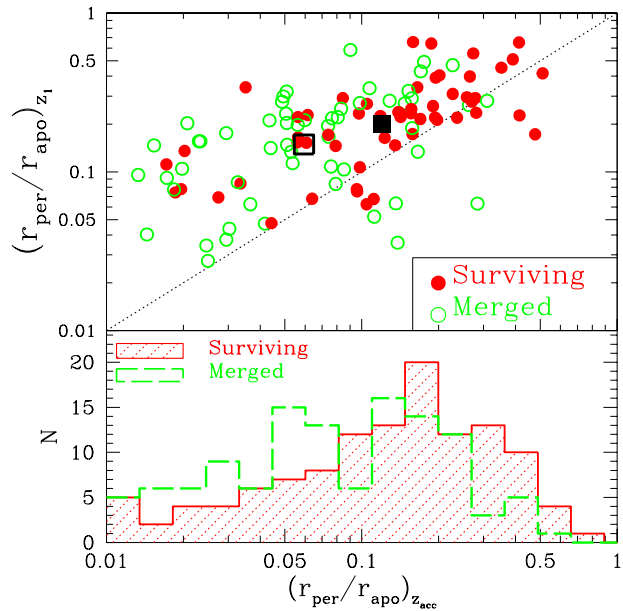}
\caption{{\it Top panel:} Orbital pericenter-to-apocenter ratio
measured at two different times during the evolution of a
satellite. Values on the horizontal axis correspond to the time of
accretion whereas values on the vertical axis are computed once
dynamical friction has eroded the apocentric distance to $\sim e^{-1}$
of its turnaround value. Most satellites lie above the $1$:$1$ dotted
line, indicating significant orbital circularization by dynamical
friction. Open and filled circles correspond, respectively, to merged
or surviving satellites at $z=0$. Open and filled squares mark the
median of each of those populations, respectively. {\it Bottom panel:}
Histogram of pericenter-to-apocenter ratio at the time of accretion
for surviving and merged satellites. Note that satellites originally
on more eccentric orbits tend to merge faster.}
\label{fig:circ}
\end{figure}
\end{center}

\subsection{Satellite evolution}

\subsubsection{Merging and survival}

Satellites are affected strongly by the tidal field of the primary,
and evolve steadily after being accreted into the halo of the host
galaxy. This is illustrated in Figure~\ref{fig:orbmass}, where the
upper panel shows the evolution of the galactocentric distance for two
satellites in one of our simulations. These two satellites follow
independent accretion paths into the halo of the primary galaxy; after
initially drifting away from the galaxy due to the universal
expansion, they reach a turnaround radius of a few hundred kpc and are
then accreted into the virial radius of the primary at similar times,
$\sim 4.5$ Gyr after the Big Bang ($z \sim 1.5$). The accretion is
indicated by the intersection between the trajectory of each satellite
in the upper panel of Figure~\ref{fig:orbmass} and the dotted line,
which shows the evolution of the virial radius of the main progenitor
of the primary.

We define the time that the satellite first enters the virial radius
of the primary as the {\it accretion time}, $t_{\rm
acc}$, or $z_{\rm acc}$, if it is expressed as a redshift. Because
masses, radii, and other characteristic properties of a satellite are
modified strongly by the tides that operate inside the halo of the
primary, it is useful to define the satellite's properties at the time
of accretion, and to refer the evolution to the values measured at
that time.

One example of the effect of tides is provided by the self-bound mass
of the satellite, whose evolution is shown in the bottom panel of
Figure~\ref{fig:orbmass}. The dark matter that remains bound to the
satellite (relative to that measured at accretion time) is shown by
open symbols; the bound mass in stars is shown by solid triangles. One
of the satellites (dashed lines) sees its orbit eroded quickly by
dynamical friction, and merges with the primary less than $4$ Gyr
after accretion, at which point the self-bound mass of the dark matter
and stellar components drops to zero. The orbital period decreases rapidly as
the satellite sinks in; we are able to trace almost 5 complete orbits
before disruption although, altogether, the satellite takes only $2.5$
Gyr to merge after the first pericentric passage, a time comparable to
just half the orbital period at accretion time.  

As the satellite is dragged inwards by dynamical friction dark matter
is lost much more readily than stars; after the first pericentric
passage only about $40\%$ of the original dark mass remains attached
to the satellite, compared with $85\%$ of the stars. This is a result
of the strong spatial segregation between stars and dark matter which
results from gas cooling and condensing at the center of dark halos
before turning into stars. Stars are only lost in large numbers at the
time of merger, when the satellite is fully disrupted by the tides.

The second satellite (solid lines in Figure~\ref{fig:orbmass})
survives as a self-bound entity until the end of the simulation. Its
orbit is affected by dynamical friction, but not as drastically as the
merged satellite: after completing 3 orbits, its apocentric distance
has only dropped from $\sim 250$ kpc at turnaround ($t_{\rm ta}\sim 3$
Gyr) to $\sim 180$ kpc at $z=0$. The stars in the satellite survive
almost unscathed; more than $85\%$ of stars remain bound to the
satellite at the end of the simulation, although only $\sim 45\%$ of the
dark matter is still attached to the satellite then.

As expected from simple dynamical friction arguments, the final fate
of a satellite regarding merging or survival depends mainly on its
mass and on the eccentricity of its orbit. The ``merged'' satellite in
Figure~\ref{fig:orbmass} is $\sim 6$ times more massive than the
``surviving'' one and is on a much more eccentric orbit: its first
pericentric radius is just $\sim 20$ kpc, compared with $45$ kpc for
the surviving satellite. More massive satellites on eccentric orbits
spiral in faster than low-mass ones, making themselves more vulnerable
to tides and full disruption.

This is confirmed in Figure~\ref{fig:orbdec}, where we show the
orbital decay timescale of all satellites identified in our
simulations as a function of their mass. Satellite masses are shown in
units of the mass of the primary galaxy at the time of accretion, and
decay timescales, $\tau$, are normalized to the orbital period of the
satellite, measured at the same time. (The timescale $\tau$ is
computed by fitting the evolution of the apocentric distance of the
satellite, a good proxy for the orbital energy, to an exponential
law.)

Surviving satellites are shown as filled circles in
Figure~\ref{fig:orbdec}, whereas open circles denote merged
satellites. More massive satellites clearly spiral in faster: $\tau$
is typically less than an orbital period for a satellite whose mass
exceeds $\sim 20\%$ of the primary. On the other hand, decay
timescales are often larger than $\sim 10$ orbital periods for
satellites with masses below $1\%$ of the primary. The dotted line
shows the $\tau \propto m^{-1}$ relation expected from simple
dynamical friction arguments \citep{binneyandtremaine87}. Most
satellites follow this trend, except perhaps for the most massive
systems, but this may just reflect difficulties estimating $\tau$ for
systems on very rapidly decaying orbits, because of poor time
sampling. The main result of these trends is a severe
underrepresentation of surviving satellites amongst massive
satellites, as shown by the distribution of satellite masses in the
bottom panel of Figure~\ref{fig:orbdec}.

\begin{center}
\begin{figure}
\includegraphics[width=84mm]{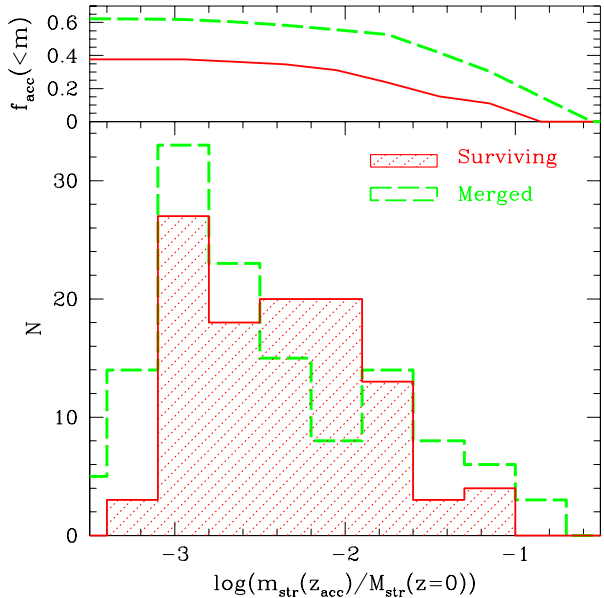}
\caption{ Distribution of satellite stellar masses measured at the
time of accretion into the host halo, and normalized to the stellar
mass of the primary at $z=0$ (bottom panel). The shaded histogram
corresponds to satellites that remain self-bound at $z=0$; the other
histogram corresponds to satellites that merge with the primary before
$z=0$. The curves in the top panel indicate the cumulative fraction of
all {\it accreted stars} contributed by each of these two
populations. Note that the ``building blocks'' of the stellar halo are
significantly more massive than the average surviving satellite.  On
average, accretion events bring about $25\%$ of the total number of
stars into the primary, $40\%$ of which remains attached to satellites
until $z=0$.  The remainder belongs to ``merged'' satellites, the
majority of which make up the stellar halo. The total number of stars
contributed by disrupted satellites exceed those locked in surviving
satellites by $\sim 50\%$.}
\label{fig:minf}
\end{figure}
\end{center}

\begin{center}
\begin{figure}
\includegraphics[width=84mm]{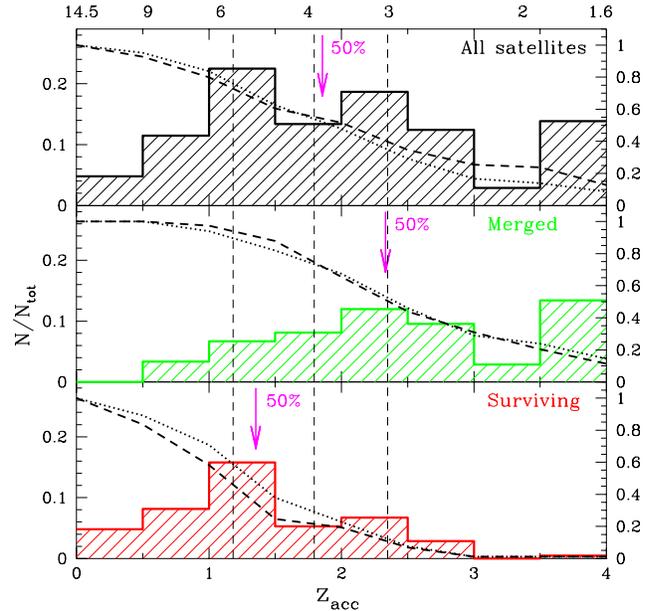}
\caption{Accretion redshift distribution of surviving (bottom panel),
merged (middle) and all (top) satellites in our simulations.  All
histograms are scaled to the total number of satellites for ease of
comparison between panels.  Dashed vertical lines indicate the
(average) redshift where the primary galaxy has accreted $25\%$,
$50\%$ and $75\%$ of its {\it total} mass at $z=0$. In each panel the
arrow shows the median satellite accretion redshift. The dotted curves
trace the cumulative distribution of satellites (by number) as a
function of $z_{\rm acc}$ (scale on right). Solid lines are like dotted
ones, but by mass.  }
\label{fig:tinf}
\end{figure}
\end{center}

\subsubsection{Orbital circularization}

As they are dragged inwards by dynamical friction, the orbital energy
of the satellites is affected more than its angular momentum and, as a
result, the satellites' orbits become gradually more circular. This is
shown in Figure~\ref{fig:circ}, where we plot the ratio between
apocentric and pericentric distance, $r_{\rm per}/r_{\rm apo}$, at the
time of accretion versus the same quantity, but measured after
dynamical friction has eroded $r_{\rm apo}$ to $e^{-1}$ of its value
at accretion.

As in Figure~\ref{fig:orbdec}, open and filled circles indicate
``merged'' and ``surviving'' satellites at $z=0$. The vast majority of
the points lie above the $1$:$1$ line, indicating that the orbits have
become significantly less eccentric with time.  Some points lie below
the dotted line, indicating the opposite effect; however, most of
these cases correspond to complex accretion where the satellite comes
as a member of a pair of satellites and is subject to three-body
interactions during accretion. (See Sales et al 2007 for further
details.)

The large open and filled squares indicate the median $r_{\rm
per}/r_{\rm apo}$ for merged and surviving satellites,
respectively. Clearly, the eccentricity of the orbit is important for
the chances of survival of a satellite: most satellites originally on
very eccentric orbits have merged with the primary by $z=0$, and the
reverse is true for surviving satellites (see bottom panel in
Figure~\ref{fig:circ}).

Satellites that merge with the primary by $z=0$ experience on average
a more substantial circularization of their orbits; the median $r_{\rm
per}/r_{\rm apo}$ evolves from $0.06$ to roughly $0.15$ in the time
it takes their orbital energies to decrease by $e^{-1}$. Further
circularization may be expected by the time that the satellite merges
with the primary and, under the right circumstances, a satellite may
even reach a nearly circular orbit before merging (see, e.g., Abadi et
al 2003b, Meza et al 2005).

Orbital circularization has been proposed as an important factor to
consider when interpreting the effects of satellite accretion events
(although see Colpi et al 1999 \nocite{colpi99} for a different 
viewpoint). Abadi et al
(2003b) argue, for example, that a satellite on a circularized orbit
might have contributed a significant fraction of the thick-disk stars
(and perhaps even some old thin-disk stars) of the Milky Way. A
further example is provided by the ``ring'' of stars discovered by the
SDSS in the anti-galactic center direction \citep{newberg02, yanny03,
helmi03}, which has been successfully modeled as
debris from the recent disruption of a satellite on a nearly circular
orbit in the outskirts of the Galactic disk \citep{penarrubia06}.
Since it is unlikely that the satellite formed on such orbit
(otherwise it would have been disrupted much earlier) its orbit has
probably evolved to become more bound and less eccentric as dynamical
friction brought the satellite nearer the Galactic disk, in agreement
with the trend shown in Figure~\ref{fig:circ}.

\begin{center}
\begin{figure}
\includegraphics[width=84mm]{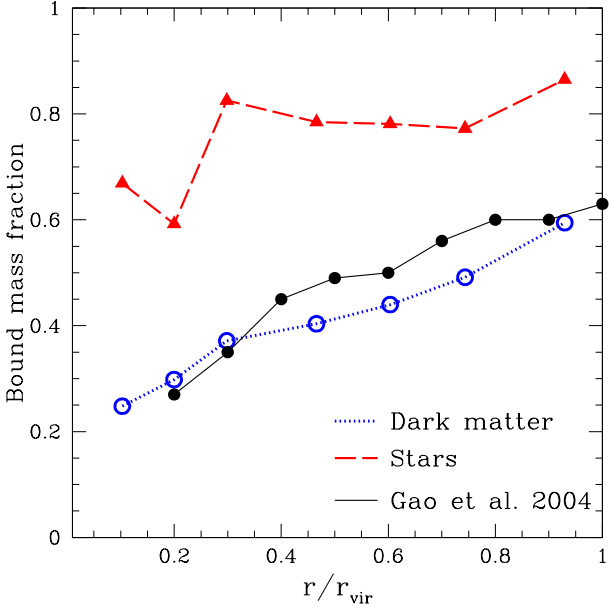}
\caption{ Mass fraction attached to surviving satellites at $z=0$,
shown as a function of radius, normalized to the virial radius of the
host. The open circles are the results of the dark matter-only
simulations of \citet{gao04b}, which are in very good agreement with
ours. This figure shows that, although surviving satellites have lost
a significant fraction of their dark mass to tides, their stellar
components have survived almost unscathed.  Overall, satellites inside
the virial radius have conserved about $40\%$ of their original dark
mass, and $\sim 75\%$ of their stars. This suggests that stars
stripped off surviving satellites are in general an unimportant
contributor to the stellar halo, and highlights the need for
simulations that include gas cooling and star formation to estimate
the importance of tidal stripping in the satellite population. }
\label{fig:mrfin}
\end{figure}
\end{center}

\subsection{Satellites and stellar halo: similarities and differences}

The main result of the trends discussed in the preceding section is
the obvious mass bias present in the population of surviving
satellites: massive satellites merge too quickly to be fairly
represented amongst satellites present at any given time. This is
shown in the bottom panel of Figure~\ref{fig:orbdec}; although the
accretion of satellites with masses exceeding $10\%$ of the host (at
the time of accretion) is not unusual, few have survived self-bound
until $z=0$.

This is also true when expressed in terms of the total stellar mass
that these accretion events have contributed to the simulated
galaxy. As shown in Figure~\ref{fig:minf}, merged satellites dominate
the high-mass end of the distribution of accreted satellites, and make
up on average $\sim 60\%$ of all accreted stars. Half of this
contribution comes in just a few massive satellites exceeding $10\%$
of the final mass in stars of the host (see upper panel in
Figure~\ref{fig:minf}). On the other hand, surviving satellites
contribute on average $\sim 40\%$ of all accreted stars and have a
combined stellar mass of about $12\%$ of the host at $z=0$. Half of
them are contributed by satellites less than $\sim 3\%$ as massive as
the host at $z=0$.

Because of the strong orbital decay dependence on mass, surviving
satellites are also biased relative to the overall population of
accreted material in terms of accretion time. This is shown
quantitatively in Figure~\ref{fig:tinf}, which shows the $z_{\rm acc}$
distribution for all satellites accreted since $z=4$ (top panel). The
bottom and middle panels, respectively, split this sample between
satellites that have either survived or merged with the host by
$z=0$. The vertical lines in this figure illustrate the average mass
accretion history of the hosts in our simulation series: from left to
right, the vertical lines indicate the average redshift when the last
$25\%$, $50\%$, and $75\%$ of the mass were assembled into the virial
radius of the host.

The accreted satellites, as a whole, trace very well this accretion
history, as may be seen from the histogram in the top panel, or by the
dotted line, which indicates the cumulative accretion history (scale
on right). Just like the total mass, half of all satellites were
accreted before $z\sim 1.8$ (see arrow labeled ``$50\%$''). The
results are quite different for ``merged satellites''; half of them
were actually accreted before $z=2.4$, which corresponds to a lookback
time of $\sim 2.7$ Gyr. Essentially no satellite accreted after $z=0.5$
has merged with the primary.  Surviving satellites, on the other hand,
are substantially biased towards late accretion. Half of them were
only accreted after $z=1.4$, and the last $25\%$ since $z \sim 1$.

Since stars brought into the galaxy by merged satellites contribute
predominantly to the stellar halo (see, e.g., Abadi et al 2006), this
result shows convincingly that substantial differences must be
expected between the stellar halo and surviving satellite population
in a galaxy built hierarchically. {\it The ``building blocks'' of the
stellar halo were on average more massive and were accreted and
disrupted much earlier than the population of satellites that survive
until the present.}

Our results provide strong support for the semianalytic modeling
results of \citet{bullockandjohnston05}. Despite the differences in
modeling techniques (these authors use theoretical merger trees to
simulate Monte Carlo accretion histories and a semianalytic approach
to dinstinguish stars and dark matter within accreted satellites), our
results agree well. For example, they find that $\sim 80\%$ of the
stellar halo is contributed by the $\sim 15$ most massive disrupted
satellites; we find, on average, $70\%$. The median accretion time for
disrupted satellites is $\sim 9$ Gyr ago; we find $\sim 10.5$ Gyr. Lastly,
they find that the median accretion time of surviving satellites was
as recently as $\sim 5$ Gyr in the past; we find $\sim 8.5$ Gyr.

As discussed by \citet{font06a,font06b}, these results may help to
explain the differences between the abundance patterns of halo stars
in the solar neighbourhood and in Galactic dwarfs \citep{fuhrmann98,
shetrone01, shetrone03, venn04}.  Although stars in both
the halo and satellites are metal-poor, the stellar halo is, at fixed
[Fe/H], more enhanced in $\alpha$ elements than stars in the dwarfs,
suggesting that its star formation and enrichment proceeded more
quickly and thoroughly than in Galactic satellites. This is
qualitatively consistent with the biases in the surviving satellite
population mentioned above. Because of the limited numerical
resolution of our simulations and our inefficient feedback recipe, we
are unable to follow accurately the metal enrichment of stars in our
simulations. Although this precludes a more detailed quantitative
comparison between simulations and observations, we regard the
distinction between satellite and stellar halo reported here as
certainly encouraging.

One final issue to consider is that, in principle, stars may also end
up in the stellar halo as a result of partial stripping of surviving
satellites. If substantial, this process might make stars in the
stellar halo difficult to differentiate from those attached to
satellites, despite the biases in mass and accretion time discussed
above. As it turns out, stripping of surviving satellites adds an
insignificant fraction of stars to the halo in our simulations; stars
stripped from surviving satellites make up a small fraction ($\sim 6\%$)
of all halo stars.

This is shown in Figure~\ref{fig:mrfin}, where we plot the fraction of
stars and dark matter that remains attached to surviving satellites as
a function of the distance to the center of the galaxy. As shown by
the filled triangles, more than $75\%$ of the stars brought into the
system by surviving satellites remain attached to them at $z=0$. We
conclude that the bulk of the halo population is not affected by stars
stripped from existing satellites, and that the substantial difference
between the stellar population of Galactic dwarfs and of the stellar
halo predicted above is robust.

\section{SUMMARY}
\label{sec:conc}

We have analyzed the properties of satellite galaxies formed in a
suite of eight N-body/gasdynamical simulations of galaxy formation in
a $\Lambda$CDM universe. Our simulations are able to resolve, at
$z=0$, the $\sim 10$ most luminous satellites orbiting around $\sim
L_*$ galaxies. We also track satellites that have merged with, or been
disrupted fully by, the primary galaxy at earlier times, giving us a
full picture of the contribution of accreted stars to the various
dynamical components of the galaxy. 

As discussed in an earlier paper of our group (Abadi et al 2006), the
stellar halo consists of stars stripped from satellites that have been
fully disrupted by the tidal field of the primary. Our analysis
here focuses on the spatial distribution, kinematics, and merging history
of the population of surviving and merged satellites, and on their
significance for the formation of the stellar halo. Our main results
may be summarized as follows.

\begin{itemize}

\item The spatial distribution of satellites at $z=0$ is consistent
with that of the dark matter in the primary galaxy's halo, and
is significantly more extended than the stellar halo. On average, half of
the $\sim 10$ brightest satellites are found within $0.37\, r_{\rm
vir}$, comparable to the half-mass radius of the dark matter
component. The half-mass radius of the stellar halo is, on the other
hand, only $0.05\, r_{\rm vir}$.

\item The kinematics of the satellite population is also similar to
the dark matter's. Satellite velocities are mildly anisotropic in the
radial direction, with $\beta_{\rm sat}\sim 0.3$-$0.4$, but not as
extreme as stars in the halo, which are found to have $\beta_{\rm
halo} \sim 0.6$-$0.8$ in the outskirts of the system. Satellite
velocity dispersions drop from the center outwards, and decrease by
about a factor of two at the virial radius from their central
value. Overall, the velocity dispersion of the satellite population is
found to provide a reasonable estimate of the halo's virial velocity:
$\sigma_{\rm sat}/V_{\rm vir} \sim 0.9 \pm 0.2$, where the
uncertainty is the rms of the eight simulations. 

\item The orbits of satellites evolve strongly after accretion as a
result of dynamical friction with the host halo and of mass stripping
by tides. More massive satellites spiral in faster than less massive
systems and are disrupted quickly as they merge with the primary,
adding their stars mainly to the stellar halo. The orbits of
satellites with masses exceeding $10\%$ of the host mass decay on
exponential timescales shorter than an orbital period, and merge
shortly after accretion. Merged satellites typically make up $\sim
63\%$ of all accreted stars in a galaxy, a substantial fraction of
which ($57\%$) was contributed by these few most massive satellites.

\item Surviving satellites are a substantially biased tracer of the
whole population of stars accreted by a galaxy. In contrast with the
``merged'' satellites that build up the halo, surviving satellites are
predominantly low-mass systems that have been accreted recently. Half
of the stars in the stellar halo were accreted before $z\sim 2.2$, and
were in satellites more massive than $\sim 6\%$ of the host at the
time of accretion. In contrast, half of the stars in surviving
satellites were brought into the system as recently as $z\sim 1.6$,
and formed in systems with masses less than $3\%$ of the host.

\item Satellite orbits are continuously circularized by dynamical
friction as they orbit within the primary's halo. The
pericenter-to-apocenter ratio typically doubles once the orbital
binding energy of the satellite has increased by a factor of $e$.

\item Stars stripped from satellites that remain self-bound until the
present make up an insignificant fraction of all stars accreted by a
galaxy, showing that, once started, the disruption process of the
stellar component of a satellite progresses on a very short
timescale. Surviving satellites conserve at $z=0$ about $75\%$ of the
stars they had at accretion time. Their surrounding dark halos, on the
other hand, have been stripped of more than $\sim 40\%$ of their mass.

\end{itemize}   

Our results offer a framework for interpreting observations of the
satellite population around luminous galaxies and for extracting
information regarding their dark matter halos. They also show that
hierarchical galaxy formation models may explain naturally the
differences in the properties of stars in the stellar halo and in
Galactic satellites highlighted by recent observational work. Although
our modeling of star formation is too simplistic (and our numerical
resolution too poor) to allow for a closer, quantitative assessment of
this issue, it is encouraging to see that, despite their differences,
stellar halos and satellites may actually be both the result of the
many accretion events that characterize galaxy formation in a
hierarchically clustering universe.

\vskip 0.5cm

\section*{Acknowledgements}
\label{acknowledgements}

LVS and MGA are grateful for the hospitality of the Max-Planck
Institute for Astrophysics in Garching, Germany, where much of the
work reported here was carried out. LVS thanks financial support from 
the Exchange of Astronomers Programme of the IAU and to the ALFA-LENAC
network. JFN acknowledges support from
Canada's NSERC, from the Leverhulme Trust, and from the Alexander von
Humboldt Foundation, as well as useful discussions with Simon White,
Alan McConnachie, and Jorge Pe\~narrubia. MS acknowledges support by
the German Science foundation (DFG) under Grant STE 710/4-1. We thank 
Scott Chapman and collaborators for sharing their results on 
Andromeda XII in advance of publication. We also thanks to the referee Andrew
Benson for useful suggestions and comments on this paper.

\bibliography{references}

\end{document}